\documentclass[a4paper,11pt]{article}

\usepackage{jheppub}
\usepackage{amsmath,amssymb,amscd}
\usepackage{graphicx}
\usepackage{url}

\graphicspath{ {fig/} }

\makeatletter
\def\hlinewd#1{%
  \noalign{\ifnum0=`}\fi\hrule \@height #1 \futurelet
   \reserved@a\@xhline}
\makeatother

\makeatletter
\renewcommand\@fpheader{}
\renewcommand\@journal{}
\makeatother

\definecolor{darkgreen}{rgb}{0.,.3,0}
\definecolor{darkblue}{rgb}{0.0,0.0,0.5}

\newcommand{\HyperInt}{\texttt{HyperInt}}
\newcommand{\Reduze}{\texttt{Reduze\;2}}

\title{
The Four-Loop Cusp Anomalous Dimension from \\
the $\mathbf{\mathcal{N} = 4}$ Sudakov Form Factor
}

\preprint{MSUHEP-19-030, P3H-19-061}

\author[a]{Tobias Huber,}
\author[b]{Andreas von Manteuffel,}
\author[c]{Erik Panzer,}
\author[b]{\\Robert M.~Schabinger,}
\author[\,d]{and Gang Yang}

\affiliation[a]{Naturwissenschaftlich-Technische Fakult\"{a}t, Universit\"{a}t Siegen,\\
Walter-Flex-Str. 3, 57068 Siegen, Germany}
\affiliation[b]{Department of Physics and Astronomy, Michigan State University,\\ East Lansing, Michigan 48824, USA}
\affiliation[c]{All Souls College, University of Oxford, OX1 4AL, Oxford, UK}
\affiliation[d]{CAS Key Laboratory of Theoretical Physics, Institute of Theoretical Physics,\\
Chinese Academy of Sciences, Beijing 100190, China}

\emailAdd{huber@physik.uni-siegen.de}
\emailAdd{vmante@msu.edu}
\emailAdd{erik.panzer@all-souls.ox.ac.uk}
\emailAdd{schabing@msu.edu}
\emailAdd{yangg@itp.ac.cn}

\abstract{
We present an analytic derivation of the full four-loop cusp anomalous dimension of $\mathcal{N}=4$ supersymmetric Yang-Mills theory from the Sudakov form factor. To extract the cusp anomalous dimension, we calculate the $\epsilon^{-2}$ pole of the form factor using parametric integrations of finite integrals.
We provide uniformly transcendental
results for the master integrals through to weight six
and confirm a very recent independent analytic result for the full four-loop cusp anomalous dimension of the $\mathcal{N}=4$ model.
}

\begin{document}
\unitlength1cm
\maketitle
\allowdisplaybreaks

\section{Introduction} 
\label{sec:intro}
The cusp anomalous dimension of $\mathcal{N} = 4$ super Yang-Mills theory has long been recognized as an important probe of the infrared structure of massless gauge theory scattering amplitudes~\cite{Korchemsky:1985xj}. In fact, the four-loop correction to the cusp anomalous dimension represents the first non-trivial check of the so-called Casimir scaling conjecture \cite{Becher:2009qa,Gardi:2009qi,Dixon:2009gx}, a proposal which would imply that the leading infrared poles of massless gauge theory scattering amplitudes have a universal, semi-classical origin. From numerical studies in $\mathcal{N} = 4$ super Yang-Mills theory \cite{Boels:2017skl,Boels:2017ftb} and QCD \cite{Moch:2017uml,Moch:2018wjh} it is now clear that the Casimir scaling conjecture must be generalized \cite{Catani:2019rvy,Becher:2019avh} to accommodate color structures built out of quartic Casimir operators. While an analytic expression for the leading-color four-loop cusp anomalous dimension in $\mathcal{N} = 4$ super Yang-Mills was proposed long ago in \cite{Bern:2006ew} (and subsequently confirmed in \cite{Beisert:2006ez,Henn:2013wfa}), the analytic sub-leading-color corrections were first discussed in the literature only recently in \cite{Henn:2019swt}.

A possible method to obtain the cusp anomalous dimension is via the calculation of the $\epsilon^{-2}$ pole of the $\mathcal{N} = 4$ Sudakov form factor. The latter was derived to three loops in~\cite{Gehrmann:2011xn}, while its four-loop integrand~\cite{Boels:2012ew}, reduction~\cite{Boels:2015yna}, transformation to a uniformly transcendental basis, and numerical integration~\cite{Boels:2017skl,Boels:2017ftb} became available over the years. The $\epsilon^{-1}$ pole of the Sudakov form factor is related to the collinear anomalous dimension, whose planar-color part is known both numerically~\cite{Cachazo:2007ad} and analytically~\cite{Dixon:2017nat}, while the non-planar-color part was computed numerically in~\cite{Boels:2017ftb}. Ongoing efforts to compute the finite (i.e.\ ${\cal O}(\epsilon^0)$) part of massless
form factors both in $\mathcal{N} = 4$ super Yang-Mills theory and QCD can be found in~\cite{Henn:2016men,vonManteuffel:2016xki,Lee:2016ixa,Lee:2017mip,vonManteuffel:2019wbj,vonManteuffel:2019gpr,Lee:2019zop,Henn:2019rmi}.

In this paper, we confirm the result of~\cite{Henn:2019swt} by providing an alternative, fully-independent analytic derivation of the full four-loop cusp anomalous dimension of the $\mathcal{N} =  4$ model.
From the $\epsilon^{-2}$ pole of the $\mathcal{N} = 4$ Sudakov form factor, we use the infrared evolution equation satisfied by the form factor \cite{Moch:2005id} to read off the cusp anomalous dimension. In order to calculate the Laurent expansions of the 55 ten-, eleven-, and twelve-line four-loop form factor master integrals which contribute to the form factor, the method of parametric integrations for a basis of finite integrals \cite{vonManteuffel:2014qoa,vonManteuffel:2015gxa} is used.
In Section \ref{sec:conventions}, we give our conventions for the Sudakov form factor of $\mathcal{N} = 4$ super Yang-Mills theory and, in Section \ref{sec:ffintegrand}, we recall the compact formula for the integrand derived in reference \cite{Boels:2017ftb}. In Section \ref{sec:ffmasters}, we present analytic results for all 55 master integrals through to $\mathcal{O}\left(\epsilon^{-2}\right)$. As expected from the analysis of \cite{Boels:2017ftb}, our results are built out of Riemann zeta values of uniform weight $8+k$ at $\mathcal{O}\left(\epsilon^{k}\right)$. We present the main results of this paper, the Laurent expansion of the Sudakov form factor and the associated cusp anomalous dimension, in Section \ref{sec:results}. Finally, we conclude in Section \ref{sec:conclusions}.

\section{Notation and Conventions}
\label{sec:conventions}
Up to an unimportant overall normalization factor, the Sudakov form factor of $\mathcal{N} = 4$ super Yang-Mills theory may be defined as in reference \cite{Gehrmann:2011xn} in terms of scalar fields,
\begin{equation}
\label{eq:def}
\mathcal{F} = \int {\rm d}^4 x \,e^{-i\, q \cdot x} \langle \phi_{1 2}^a(p_1) \phi_{1 2}^b(p_2) | \left[\phi_{3 4}^c \phi_{3 4}^c\right]\!(x) \,|0\rangle.
\end{equation}
In Eq. \eqref{eq:def}, field superscripts denote adjoint $SU(N_c)$ color indices and field subscripts denote $SU(4)_R$ indices. It is convenient to expand the Sudakov form factor in a modified bare 't Hooft coupling,
\begin{equation}
g^2 = \frac{N_c\, g_{\scriptscriptstyle \mathcal{N} = 4}^2}{16\pi^2}\left(4\pi e^{-\gamma_{ \mathrm{E}}}\right)^\epsilon,
\end{equation}
where $N_c$ is the number of colors, $g_{\scriptscriptstyle \mathcal{N} = 4}$ is the bare coupling of the model, $\gamma_{ \mathrm E}$ is Euler's constant, and $\epsilon = (4-d)/2$ is the parameter of dimensional regularization. At each order in perturbation theory, the perturbative expansion coefficients of the Sudakov form factor depend on a virtuality parameter, $q^2 = (p_1 + p_2)^2$, which may be set to $-1$ without loss of generality. In terms of the modified bare 't Hooft coupling, we then have
\begin{equation}
 \mathcal{F} = \mathcal{F}^{tree} \sum_{\ell = 0}^\infty g^{2\ell}F^{(\ell)}
\end{equation}
for the Sudakov form factor.

\section{Integrand for the $\mathbf{\mathcal{N} = 4}$ Four-Loop Sudakov Form Factor}
\label{sec:ffintegrand}
In this section, we recall the compact form presented in Section 4 of reference \cite{Boels:2017ftb} for the unintegrated four-loop Sudakov form factor. The expression below was derived in earlier work using loop-level color-kinematics duality \cite{Boels:2012ew}, integration by parts identities \cite{Boels:2015yna,Tkachov:1981wb,Chetyrkin:1981qh,Laporta:2001dd}, and a systematic construction of and projection onto (conjectured) uniform-transcendentality master integrals \cite{Boels:2017ftb}. As expected, it splits naturally into a leading-color (planar-color) part and a sub-leading-color (non-planar-color) part.\footnote{Note that non-trivial non-planar master integrals also contribute to the leading-color part.} We have
\begin{align}
\label{eq:integrand}
 &F^{(4)} = 2\Bigg[8 I_{\mathrm{p}, 1}^{(1)}+2 I_{\mathrm{p}, 2}^{(2)}-2 I_{\mathrm{p}, 3}^{(3)}+2 I_{\mathrm{p}, 4}^{(4)}+\frac{1}{2} I_{\mathrm{p}, 5}^{(5)}+2 I_{\mathrm{p}, 6}^{(6)}+4 I_{\mathrm{p}, 7}^{(7)}+2 I_{\mathrm{p}, 8}^{(9)}-2 I_{\mathrm{p}, 9}^{(10)}+ I_{\mathrm{p}, 10}^{(12)}
 \nonumber \\
 &+ I_{\mathrm{p}, 11}^{(12)}+2 I_{\mathrm{p}, 12}^{(13)}+2 I_{\mathrm{p}, 13}^{(14)}-2 I_{\mathrm{p}, 14}^{(17)}+2 I_{\mathrm{p}, 15}^{(17)}-2 I_{\mathrm{p}, 16}^{(19)}+ I_{\mathrm{p}, 17}^{(19)}+ I_{\mathrm{p}, 18}^{(21)}+\frac{1}{2} I_{\mathrm{p}, 19}^{(25)}+2 I_{\mathrm{p}, 20}^{(30)}+2 I_{\mathrm{p}, 21}^{(13)}
 \nonumber \\
 &+4 I_{\mathrm{p}, 22}^{(14)}-2 I_{\mathrm{p}, 23}^{(14)}- I_{\mathrm{p}, 24}^{(14)}+4 I_{\mathrm{p}, 25}^{(17)}- I_{\mathrm{p}, 26}^{(17)}-2 I_{\mathrm{p}, 27}^{(17)}-2 I_{\mathrm{p}, 28}^{(17)}- I_{\mathrm{p}, 29}^{(19)}
- I_{\mathrm{p}, 30}^{(19)}+ I_{\mathrm{p}, 31}^{(19)}-\frac{1}{2} I_{\mathrm{p}, 32}^{(30)}\Bigg]
 \nonumber \\
 &+\frac{48}{N_c^2}\Bigg[ \frac{1}{2} I_{1}^{(21)}+\frac{1}{2} I_{2}^{(22)}+\frac{1}{2} I_{3}^{(23)}- I_{4}^{(24)}+\frac{1}{4} I_{5}^{(25)}-\frac{1}{4} I_{6}^{(26)}-\frac{1}{4} I_{7}^{(26)}+2 I_{8}^{(27)}+I_{9}^{(28)}
 \nonumber \\
 &+4 I_{10}^{(29)}+ I_{11}^{(30)}+ I_{12}^{(27)}-\frac{1}{2} I_{13}^{(28)}+ I_{14}^{(29)}+ I_{15}^{(29)}+ I_{16}^{(30)}+ I_{17}^{(30)}+ I_{18}^{(30)}+ I_{19}^{(22)}+ I_{20}^{(22)}
 \nonumber \\
 &- I_{21}^{(24)}+\frac{1}{4} I_{22}^{(24)}+\frac{1}{2} I_{23}^{(28)}\Bigg],
\end{align}
where $F^{(4)}$ is defined in Section \ref{sec:conventions}.
The precise definition of the integrals in Eq.~\eqref{eq:integrand} can be found in \cite{Boels:2017ftb}.
Note also that three integrals appear in both the planar-color and non-planar-color parts:
\begin{equation}
I_{1}^{(21)} = I_{\mathrm{p}, 18}^{(21)} , \qquad 
I_{5}^{(25)} = I_{\mathrm{p}, 19}^{(25)} , \qquad
I_{11}^{(30)} = I_{\mathrm{p}, 20}^{(30)}.
\end{equation}
Laurent expansions of the master integrals through to $\mathcal{O}(\epsilon^{-2})$ are provided in the next section.

\section{Master Integrals to Weight Six}
\label{sec:ffmasters}
In this section, we provide results for the (conjectured) uniform-transcendentality master integrals which appear on the right-hand side of Eq. \eqref{eq:integrand}, up to and including terms of transcendental weight six.
From the definitions of Section \ref{sec:conventions}, it is evident that we employ the $\overline{\rm MS}$ normalization convention for our master integrals. In total, 32 integrals contribute to the planar-color part and 23 integrals contribute to the non-planar-color part.
\begin{align}
\label{eq:masters}
I_{\mathrm{p}, 1}^{(1)} &=\frac{1}{\epsilon ^8}\left(\frac{1}{576}\right)+\frac{1}{\epsilon ^6}\left(\frac{17}{288} \zeta _2\right)+\frac{1}{\epsilon ^5}\left(\frac{89}{432} \zeta _3\right)+\frac{1}{\epsilon ^4}\left(\frac{677}{720}\zeta _2^2\right)
\nonumber \\
&+\frac{1}{\epsilon ^3}\left(\frac{5489}{720} \zeta_5+\frac{487}{216} \zeta _3 \zeta _2\right)+\frac{1}{\epsilon ^2}\left(\frac{1571}{324} \zeta _3^2+\frac{3919}{420} \zeta _2^3\right)+\mathcal{O}\left(\epsilon^{-1}\right)
\\
I_{\mathrm{p}, 2}^{(2)} &= \frac{1}{\epsilon ^8}\left(\frac{1}{144}\right)+\frac{1}{\epsilon ^6}\left(-\frac{13}{144} \zeta _2\right)+\frac{1}{\epsilon ^5}\left(-\frac{577}{432}\zeta _3\right)+\frac{1}{\epsilon ^4}\left(-\frac{269}{80}\zeta _2^2\right)
\nonumber \\
&+\frac{1}{\epsilon ^3}\left(-\frac{4309}{720} \zeta_5-\frac{236}{27} \zeta_3 \zeta_2\right)+\frac{1}{\epsilon ^2}\left(\frac{115529}{1296} \zeta_3^2-\frac{13721}{1260} \zeta_2^3\right)+\mathcal{O}\left(\epsilon^{-1}\right)
\\
I_{\mathrm{p}, 3}^{(3)} &= \frac{1}{\epsilon ^8}\left(-\frac{1}{288}\right)+\frac{1}{\epsilon ^6}\left(-\frac{17}{144} \zeta _2\right)+\frac{1}{\epsilon ^5}\left(-\frac{233}{216} \zeta _3\right)+\frac{1}{\epsilon ^4}\left(-\frac{173}{360}\zeta _2^2\right)
\nonumber \\
&+\frac{1}{\epsilon ^3}\left(-\frac{16529}{360}\zeta
   _5+\frac{2033}{108} \zeta _3 \zeta _2\right)+\frac{1}{\epsilon ^2}\left(-\frac{8717}{162} \zeta _3^2-\frac{615}{14}\zeta _2^3\right)+\mathcal{O}\left(\epsilon^{-1}\right)
\\
I_{\mathrm{p}, 4}^{(4)} &= \frac{1}{\epsilon ^8}\left(\frac{1}{288}\right)+\frac{1}{\epsilon ^6}\left(\frac{17}{144} \zeta _2\right)+\frac{1}{\epsilon ^5}\left(\frac{89}{216}\zeta _3\right)+\frac{1}{\epsilon ^4}\left(\frac{533}{360} \zeta _2^2\right)
\nonumber \\
&+\frac{1}{\epsilon ^3}\left(\frac{7469}{360} \zeta
   _5+\frac{163}{108} \zeta _3 \zeta _2\right)+\frac{1}{\epsilon ^2}\left(\frac{1150}{81} \zeta _3^2+\frac{218}{35} \zeta _2^3\right)+\mathcal{O}\left(\epsilon^{-1}\right)
\\
I_{\mathrm{p}, 5}^{(5)} &= \frac{1}{\epsilon ^8}\left(\frac{1}{72}\right)+\frac{1}{\epsilon ^6}\left(-\frac{13}{72} \zeta _2\right)+\frac{1}{\epsilon ^5}\left(-\frac{577}{216} \zeta _3\right)+\frac{1}{\epsilon ^4}\left(-\frac{887}{120} \zeta _2^2\right)
\nonumber \\
&+\frac{1}{\epsilon ^3}\left(-\frac{21109}{360} \zeta
   _5-\frac{4}{27} \zeta _3 \zeta _2\right)+\frac{1}{\epsilon ^2}\left(\frac{193721}{648} \zeta _3^2-\frac{2897}{30}\zeta_2^3\right)+\mathcal{O}\left(\epsilon^{-1}\right)
\\
I_{\mathrm{p}, 6}^{(6)} &= \frac{1}{\epsilon ^8}\left(\frac{1}{576}\right)+\frac{1}{\epsilon ^6}\left(\frac{7}{144} \zeta _2\right)+\frac{1}{\epsilon ^5}\left(\frac{169}{864} \zeta _3\right)+ \frac{1}{\epsilon ^4}\left(\frac{713}{1440} \zeta _2^2\right)
\nonumber \\
&+\frac{1}{\epsilon ^3}\left(\frac{3013}{1440} \zeta_5+\frac{115}{216} \zeta _3 \zeta _2\right)+\frac{1}{\epsilon ^2}\left(-\frac{13919}{2592}\zeta _3^2+\frac{1759}{7560} \zeta _2^3\right)+\mathcal{O}\left(\epsilon^{-1}\right)
\\
I_{\mathrm{p}, 7}^{(7)} &= \frac{1}{\epsilon ^8}\left(\frac{11}{576}\right)+\frac{1}{\epsilon ^6}\left(\frac{11}{48} \zeta _2\right)+\frac{1}{\epsilon ^5}\left(\frac{1937}{864} \zeta _3\right)+\frac{1}{\epsilon ^4}\left(\frac{487}{360} \zeta _2^2\right)
\nonumber \\
&+\frac{1}{\epsilon ^3}\left(\frac{94313}{1440} \zeta _5-\frac{1505}{48} \zeta _3 \zeta_2\right)+\frac{1}{\epsilon ^2}\left(-\frac{14483}{324}\zeta _3^2+\frac{35053}{1260} \zeta _2^3\right)+\mathcal{O}\left(\epsilon^{-1}\right)
\\
I_{\mathrm{p}, 8}^{(9)} &= \frac{1}{\epsilon ^8}\left(\frac{1}{576}\right)+\frac{1}{\epsilon ^6}\left(\frac{1}{24}\zeta _2\right)+\frac{1}{\epsilon ^5}\left(\frac{163}{864} \zeta _3\right)+\frac{1}{\epsilon ^4}\left(\frac{161}{160} \zeta _2^2\right)+\frac{1}{\epsilon ^3}\left(\frac{5803}{1440}\zeta_5+\frac{253}{36}\zeta _3 \zeta _2\right)
\nonumber \\
&+\frac{1}{\epsilon ^2}\left(\frac{59509}{2592} \zeta _3^2+\frac{119}{6} \zeta _2^3\right)+\mathcal{O}\left(\epsilon^{-1}\right)
\\
I_{\mathrm{p}, 9}^{(10)} &= \frac{1}{\epsilon ^8}\left(-\frac{13}{576}\right)+\frac{1}{\epsilon ^6}\left(\frac{5}{48} \zeta _2\right)+\frac{1}{\epsilon ^5}\left(\frac{743}{864} \zeta _3\right)+\frac{1}{\epsilon ^4}\left(\frac{167}{480} \zeta _2^2\right)
\nonumber \\
&+\frac{1}{\epsilon ^3}\left(\frac{82931}{1440} \zeta _5-\frac{179}{9} \zeta _3 \zeta_2\right)+\frac{1}{\epsilon ^2}\left(\frac{425345}{2592} \zeta _3^2+\frac{4163}{90} \zeta _2^3\right)+\mathcal{O}\left(\epsilon^{-1}\right)
\\
I_{\mathrm{p}, 10}^{(12)} &= \frac{1}{\epsilon ^8}\left(-\frac{1}{72}\right)+\frac{1}{\epsilon ^6}\left(-\frac{29}{144} \zeta _2\right)+\frac{1}{\epsilon ^5}\left(-\frac{577}{432} \zeta _3\right)+\frac{1}{\epsilon ^4}\left(\frac{31}{240} \zeta _2^2\right)
\nonumber \\
&+\frac{1}{\epsilon ^3}\left(-\frac{36367}{720} \zeta
   _5+\frac{4019}{108} \zeta _3 \zeta _2\right)+\frac{1}{\epsilon ^2}\left(\frac{128729}{1296} \zeta _3^2-\frac{4741}{252} \zeta _2^3\right)+\mathcal{O}\left(\epsilon^{-1}\right)
\\
I_{\mathrm{p}, 11}^{(12)} &= \frac{1}{\epsilon ^8}\left(\frac{1}{144}\right)+\frac{1}{\epsilon ^6}\left(-\frac{1}{144}\zeta _2\right)+\frac{1}{\epsilon ^5}\left(-\frac{1}{432}\zeta _3\right)+\frac{1}{\epsilon ^4}\left(\frac{57}{80} \zeta _2^2\right)
\nonumber \\
&+\frac{1}{\epsilon^3}\left(-\frac{38149}{720} \zeta _5+\frac{1061}{54} \zeta _3 \zeta _2\right)+\frac{1}{\epsilon ^2}\left(-\frac{237775}{1296} \zeta _3^2-\frac{37363}{2520} \zeta _2^3\right)+\mathcal{O}\left(\epsilon^{-1}\right)
\\
I_{\mathrm{p}, 12}^{(13)} &= \frac{1}{\epsilon ^8}\left(\frac{1}{576}\right)+\frac{1}{\epsilon ^6}\left(\frac{1}{24}\zeta _2\right)+\frac{1}{\epsilon ^5}\left(\frac{181}{864} \zeta _3\right)+\frac{1}{\epsilon ^4}\left(\frac{57}{80}\zeta _2^2\right)+\frac{1}{\epsilon
   ^3}\left(\frac{5833}{1440} \zeta _5+\frac{595}{144}\zeta _3 \zeta _2\right)
\nonumber \\
&+\frac{1}{\epsilon ^2}\left(\frac{2083}{162} \zeta _3^2+\frac{33163}{2520} \zeta _2^3\right)+\mathcal{O}\left(\epsilon^{-1}\right)
\\
I_{\mathrm{p}, 13}^{(14)} &= \frac{1}{\epsilon ^8}\left(\frac{23}{576}\right)+\frac{1}{\epsilon ^6}\left(-\frac{47}{144} \zeta _2\right)+\frac{1}{\epsilon ^5}\left(-\frac{1789}{864} \zeta _3\right)+\frac{1}{\epsilon ^4}\left(-\frac{433}{288} \zeta _2^2\right)
\nonumber \\
&+\frac{1}{\epsilon ^3}\left(-\frac{60961}{1440} \zeta_5+\frac{4765}{216} \zeta _3 \zeta _2\right)+\frac{1}{\epsilon ^2}\left(\frac{134567}{2592} \zeta _3^2-\frac{52957}{2520} \zeta _2^3\right)+\mathcal{O}\left(\epsilon^{-1}\right)
\\
I_{\mathrm{p}, 14}^{(17)} &= \frac{1}{\epsilon ^8}\left(-\frac{3}{64}\right)+\frac{1}{\epsilon ^6}\left(\frac{31}{96} \zeta _2\right)+\frac{1}{\epsilon ^5}\left(\frac{3}{4} \zeta _3\right)+\frac{1}{\epsilon ^4}\left(-\frac{6541}{1440} \zeta _2^2\right)
\nonumber \\
&+\frac{1}{\epsilon ^3}\left(-\frac{1063}{20} \zeta
   _5-\frac{781}{36} \zeta _3 \zeta _2\right)+\frac{1}{\epsilon ^2}\left(\frac{2741}{144} \zeta _3^2-\frac{192937}{2016} \zeta _2^3\right)+\mathcal{O}\left(\epsilon^{-1}\right)
\\
I_{\mathrm{p}, 15}^{(17)} &= \frac{1}{\epsilon ^8}\left(\frac{1}{576}\right)+\frac{1}{\epsilon ^6}\left(-\frac{1}{8}\zeta _2\right)+\frac{1}{\epsilon ^5}\left(-\frac{319}{432} \zeta _3\right)+\frac{1}{\epsilon ^4}\left(\frac{1201}{2880} \zeta _2^2\right)
\nonumber \\
&+\frac{1}{\epsilon ^3}\left(-\frac{1373}{360} \zeta
   _5+\frac{2353}{144} \zeta _3 \zeta _2\right)+\frac{1}{\epsilon ^2}\left(\frac{7328}{81}\zeta _3^2+\frac{67729}{4032} \zeta _2^3\right)+\mathcal{O}\left(\epsilon^{-1}\right)
\\
I_{\mathrm{p}, 16}^{(19)} &= \frac{1}{\epsilon ^8}\left(-\frac{1}{96}\right)+\frac{1}{\epsilon ^6}\left(-\frac{53}{288} \zeta _2\right)+\frac{1}{\epsilon ^5}\left(-\frac{337}{288} \zeta _3\right)+\frac{1}{\epsilon ^4}\left(-\frac{637}{240} \zeta _2^2\right)
\nonumber \\
&+\frac{1}{\epsilon ^3}\left(-\frac{27601}{480} \zeta_5+\frac{1541}{216} \zeta _3 \zeta _2\right)+\frac{1}{\epsilon ^2}\left(-\frac{4069}{288} \zeta _3^2-\frac{524371}{10080} \zeta _2^3\right)+\mathcal{O}\left(\epsilon^{-1}\right)
\\
I_{\mathrm{p}, 17}^{(19)} &= \frac{1}{\epsilon ^8}\left(\frac{5}{288}\right)+\frac{1}{\epsilon ^6}\left(-\frac{139}{288} \zeta _2\right)+\frac{1}{\epsilon ^5}\left(-\frac{3809}{864}\zeta _3\right)+\frac{1}{\epsilon ^4}\left(-\frac{2609}{240} \zeta _2^2\right)
\nonumber \\
&+\frac{1}{\epsilon ^3}\left(-\frac{56425}{288} \zeta
   _5+\frac{10235}{432} \zeta _3 \zeta _2\right)+\frac{1}{\epsilon ^2}\left(\frac{80651}{1296} \zeta _3^2-\frac{184073}{1008}\zeta _2^3\right)+\mathcal{O}\left(\epsilon^{-1}\right)
\\
I_{\mathrm{p}, 18}^{(21)} &= \frac{1}{\epsilon ^8}\left(\frac{1}{576}\right)+\frac{1}{\epsilon ^6}\left(\frac{1}{36}\zeta _2\right)+\frac{1}{\epsilon ^5}\left(\frac{151}{864} \zeta _3\right)+\frac{1}{\epsilon ^4}\left(\frac{173}{288} \zeta _2^2\right)+\frac{1}{\epsilon ^3}\left(\frac{5503}{1440} \zeta
   _5+\frac{505}{216} \zeta _3 \zeta _2\right)
\nonumber \\
&+\frac{1}{\epsilon ^2}\left(\frac{9895}{2592} \zeta _3^2+\frac{6317}{720} \zeta _2^3\right)+\mathcal{O}\left(\epsilon^{-1}\right)
\\
I_{\mathrm{p}, 19}^{(25)} &= \frac{1}{\epsilon ^8}\left(\frac{1}{288}\right)+\frac{1}{\epsilon ^6}\left(\frac{1}{144}\zeta _2\right)+\frac{1}{\epsilon ^5}\left(\frac{209}{216} \zeta _3\right)+\frac{1}{\epsilon ^4}\left(\frac{623}{120} \zeta _2^2\right)
\nonumber \\
&+\frac{1}{\epsilon ^3}\left(\frac{39449}{360} \zeta _5-\frac{205}{108} \zeta _3 \zeta
   _2\right)+\frac{1}{\epsilon ^2}\left(\frac{11621}{162}\zeta _3^2+\frac{38501}{315} \zeta _2^3\right)+\mathcal{O}\left(\epsilon^{-1}\right)
\\
I_{\mathrm{p}, 20}^{(30)} &= \frac{1}{\epsilon ^8}\left(\frac{1}{288}\right)+\frac{1}{\epsilon ^6}\left(-\frac{1}{32}\zeta _2\right)+\frac{1}{\epsilon ^5}\left(-\frac{187}{864} \zeta _3\right)+\frac{1}{\epsilon ^4}\left(-\frac{403}{720} \zeta _2^2\right)
\nonumber \\
&+\frac{1}{\epsilon ^3}\left(-\frac{38659}{1440} \zeta
   _5+\frac{191}{36} \zeta _3 \zeta _2\right)+\frac{1}{\epsilon ^2}\left(-\frac{14047}{2592}\zeta _3^2-\frac{284189}{10080} \zeta _2^3\right)+\mathcal{O}\left(\epsilon^{-1}\right)
\\
I_{\mathrm{p}, 21}^{(13)} &= \frac{1}{\epsilon ^5}\left(\frac{1}{24}\zeta _3\right)+\frac{1}{\epsilon ^3}\left(\frac{7}{12} \zeta _5-\frac{5}{12} \zeta _3 \zeta _2\right)+\frac{1}{\epsilon ^2}\left(-\frac{193}{72} \zeta _3^2+\frac{6389}{2520} \zeta _2^3\right)+\mathcal{O}\left(\epsilon^{-1}\right)
\\
I_{\mathrm{p}, 22}^{(14)} &= \frac{1}{\epsilon ^6}\left(\frac{1}{48}\zeta _2\right)+\frac{1}{\epsilon ^5}\left(-\frac{7}{48} \zeta _3\right)+\frac{1}{\epsilon ^4}\left(-\frac{13}{240} \zeta _2^2\right)+\frac{1}{\epsilon
   ^3}\left(-\frac{281}{48} \zeta _5+\frac{17}{9}\zeta _3 \zeta _2\right)
\nonumber \\
&+\frac{1}{\epsilon ^2}\left(\frac{439}{144} \zeta _3^2-\frac{8053}{2520} \zeta _2^3\right)+\mathcal{O}\left(\epsilon^{-1}\right)
\\
I_{\mathrm{p}, 23}^{(14)} &= \frac{1}{\epsilon ^4}\left(-\frac{7}{20} \zeta _2^2\right)+\frac{1}{\epsilon ^3}\left(-\frac{377}{24} \zeta _5+\frac{97}{12} \zeta _3 \zeta _2\right)+\frac{1}{\epsilon
   ^2}\left(\frac{433}{24} \zeta _3^2-\frac{8531}{840} \zeta _2^3\right)+\mathcal{O}\left(\epsilon^{-1}\right)
\\
I_{\mathrm{p}, 24}^{(14)} &= \frac{1}{\epsilon ^8}\left(\frac{5}{48}\right)+\frac{1}{\epsilon ^6}\left(-\frac{65}{72} \zeta _2\right)+\frac{1}{\epsilon ^5}\left(-\frac{293}{48} \zeta _3\right)+\frac{1}{\epsilon ^4}\left(-\frac{2171}{480} \zeta _2^2\right)
\nonumber \\
&+\frac{1}{\epsilon ^3}\left(-\frac{4019}{48} \zeta
   _5+\frac{11495}{216} \zeta _3 \zeta _2\right)+\frac{1}{\epsilon ^2}\left(\frac{82361}{432} \zeta _3^2-\frac{163871}{10080}\zeta _2^3\right)+\mathcal{O}\left(\epsilon^{-1}\right)
\\
I_{\mathrm{p}, 25}^{(17)} &= \frac{1}{\epsilon ^8}\left(\frac{1}{96}\right)+\frac{1}{\epsilon ^6}\left(-\frac{11}{96} \zeta _2\right)+\frac{1}{\epsilon ^5}\left(-\frac{11}{9} \zeta _3\right)+\frac{1}{\epsilon ^4}\left(-\frac{2743}{960} \zeta _2^2\right)
\nonumber \\
&+\frac{1}{\epsilon ^3}\left(-\frac{2329}{80} \zeta
   _5-\frac{11}{36} \zeta _3 \zeta _2\right)+\frac{1}{\epsilon ^2}\left(\frac{26141}{864} \zeta _3^2-\frac{642007}{20160}\zeta _2^3\right)+\mathcal{O}\left(\epsilon^{-1}\right)
\\
I_{\mathrm{p}, 26}^{(17)} &= \frac{1}{\epsilon ^8}\left(\frac{5}{144}\right)+\frac{1}{\epsilon ^6}\left(-\frac{2}{9} \zeta _2\right)+\frac{1}{\epsilon ^5}\left(-\frac{331}{216} \zeta _3\right)+\frac{1}{\epsilon ^4}\left(\frac{1171}{240} \zeta _2^2\right)
\nonumber \\
&+\frac{1}{\epsilon ^3}\left(\frac{3857}{36} \zeta
   _5+\frac{2041}{216} \zeta _3 \zeta _2\right)+\frac{1}{\epsilon ^2}\left(\frac{72223}{1296} \zeta _3^2+\frac{67171}{504} \zeta _2^3\right)+\mathcal{O}\left(\epsilon^{-1}\right)
\\
I_{\mathrm{p}, 27}^{(17)} &= \frac{1}{\epsilon ^5}\left(\frac{1}{48}\zeta _3\right)+\frac{1}{\epsilon ^4}\left(-\frac{7}{160} \zeta _2^2\right)+\frac{1}{\epsilon ^3}\left(\frac{91}{48} \zeta _5-\frac{113}{48}\zeta _3 \zeta _2\right)
\nonumber \\   
&+\frac{1}{\epsilon ^2}\left(\frac{1063}{288}\zeta
   _3^2-\frac{13751}{5040} \zeta _2^3\right)+\mathcal{O}\left(\epsilon^{-1}\right)
\\
I_{\mathrm{p}, 28}^{(17)} &= \frac{1}{\epsilon ^6}\left(\frac{1}{24}\zeta _2\right)+\frac{1}{\epsilon ^5}\left(-\frac{1}{8}\zeta _3\right)+\frac{1}{\epsilon ^4}\left(\frac{1}{60}\zeta _2^2\right)+\frac{1}{\epsilon
   ^3}\left(\frac{20}{3} \zeta _5-\frac{29}{36} \zeta _3 \zeta _2\right)
\nonumber \\
&+\frac{1}{\epsilon ^2}\left(\frac{703}{24} \zeta _3^2+\frac{11281}{1260} \zeta _2^3\right)+\mathcal{O}\left(\epsilon^{-1}\right)
\\
I_{\mathrm{p}, 29}^{(19)} &= \frac{1}{\epsilon ^8}\left(-\frac{1}{36}\right)+\frac{1}{\epsilon ^6}\left(-\frac{1}{16}\zeta _2\right)+\frac{1}{\epsilon ^5}\left(-\frac{107}{432} \zeta _3\right)+\frac{1}{\epsilon ^4}\left(\frac{9}{2} \zeta _2^2\right)
\nonumber \\
&+\frac{1}{\epsilon
   ^3}\left(\frac{18091}{720} \zeta _5+\frac{467}{12} \zeta _3 \zeta _2\right)+\frac{1}{\epsilon ^2}\left(\frac{155179}{1296}\zeta _3^2+\frac{348347}{5040}\zeta _2^3\right)+\mathcal{O}\left(\epsilon^{-1}\right)
\\
I_{\mathrm{p}, 30}^{(19)} &= \frac{1}{\epsilon ^5}\left(-\frac{7}{24} \zeta _3\right)+\frac{1}{\epsilon ^4}\left(-\frac{5}{48} \zeta _2^2\right)+\frac{1}{\epsilon ^3}\left(\frac{69}{8} \zeta _5+\frac{7}{4} \zeta _3 \zeta _2\right)
\nonumber \\
&+\frac{1}{\epsilon ^2}\left(\frac{1885}{72} \zeta _3^2+\frac{8131}{1008} \zeta_2^3\right)+\mathcal{O}\left(\epsilon^{-1}\right)
\\
I_{\mathrm{p}, 31}^{(19)} &= \frac{1}{\epsilon ^8}\left(-\frac{11}{288}\right)+\frac{1}{\epsilon ^6}\left(\frac{65}{288} \zeta _2\right)+\frac{1}{\epsilon ^5}\left(-\frac{2005}{864} \zeta _3\right)+\frac{1}{\epsilon ^4}\left(\frac{63}{80} \zeta _2^2\right)
\nonumber \\
&+\frac{1}{\epsilon ^3}\left(\frac{115559}{1440} \zeta _5-\frac{4519}{432} \zeta _3 \zeta
   _2\right)+\frac{1}{\epsilon ^2}\left(\frac{18203}{162}\zeta _3^2+\frac{22915}{252} \zeta _2^3\right)+\mathcal{O}\left(\epsilon^{-1}\right)
\\
I_{\mathrm{p}, 32}^{(30)} &= \frac{1}{\epsilon ^8}\left(-\frac{1}{12}\right)+\frac{1}{\epsilon ^6}\left(\frac{35}{48} \zeta _2\right)+\frac{1}{\epsilon ^5}\left(\frac{445}{144} \zeta _3\right)+\frac{1}{\epsilon ^4}\left(-\frac{269}{240} \zeta _2^2\right)
\nonumber \\
&+\frac{1}{\epsilon ^3}\left(\frac{2767}{80} \zeta _5-\frac{1433}{36}\zeta_3 \zeta_2\right)+\frac{1}{\epsilon ^2}\left(-\frac{14051}{432}\zeta _3^2-\frac{8363}{630} \zeta _2^3\right)+\mathcal{O}\left(\epsilon^{-1}\right)
\\
I_{1}^{(21)} &= I_{\mathrm{p}, 18}^{(21)}
\\
I_{2}^{(22)} &= \frac{1}{\epsilon ^8}\left(\frac{1}{192}\right)+\frac{1}{\epsilon ^6}\left(-\frac{19}{72}\zeta _2\right)+\frac{1}{\epsilon ^5}\left(-\frac{61}{32} \zeta _3\right)+\frac{1}{\epsilon ^4}\left(-\frac{5089}{1440} \zeta _2^2\right)
\nonumber \\
&+\frac{1}{\epsilon ^3}\left(-\frac{41237}{480}\zeta
   _5+\frac{4111}{216} \zeta _3 \zeta _2\right)+\frac{1}{\epsilon ^2}\left(-\frac{2881}{864} \zeta _3^2-\frac{8259}{112} \zeta _2^3\right)+\mathcal{O}\left(\epsilon^{-1}\right)
\\
I_{3}^{(23)} &= \frac{1}{\epsilon ^8}\left(\frac{1}{144}\right)+\frac{1}{\epsilon ^6}\left(-\frac{5}{18} \zeta _2\right)+\frac{1}{\epsilon ^5}\left(-\frac{401}{216} \zeta _3\right)+\frac{1}{\epsilon ^4}\left(\frac{19}{16} \zeta _2^2\right)
\nonumber \\
&+\frac{1}{\epsilon ^3}\left(-\frac{16277}{360} \zeta
   _5+\frac{13151}{216} \zeta _3 \zeta _2\right)+\frac{1}{\epsilon ^2}\left(\frac{248513}{1296} \zeta _3^2+\frac{751}{45} \zeta _2^3\right)+\mathcal{O}\left(\epsilon^{-1}\right)
\\
I_{4}^{(24)} &= \frac{1}{\epsilon ^8}\left(-\frac{5}{576}\right)+\frac{1}{\epsilon ^6}\left(\frac{65}{144} \zeta _2\right)+\frac{1}{\epsilon ^5}\left(\frac{1645}{864}\zeta _3\right)+\frac{1}{\epsilon ^4}\left(-\frac{109}{40} \zeta _2^2\right)
\nonumber \\
&+\frac{1}{\epsilon ^3}\left(\frac{2093}{288} \zeta _5-\frac{9361}{216} \zeta _3 \zeta
   _2\right)+\frac{1}{\epsilon ^2}\left(-\frac{166229}{2592}\zeta _3^2-\frac{289223}{10080} \zeta _2^3\right)+\mathcal{O}\left(\epsilon^{-1}\right)
\\
I_{5}^{(25)} &= I_{\mathrm{p}, 19}^{(25)}
\\
I_{6}^{(26)} &= \frac{1}{\epsilon^8}\left(-\frac{25}{576}\right)+\frac{1}{\epsilon ^6}\left(\frac{313}{288} \zeta _2\right)+\frac{1}{\epsilon ^5}\left(\frac{1241}{216} \zeta _3\right)+\frac{1}{\epsilon ^4}\left(-\frac{3671}{720} \zeta _2^2\right)
\nonumber \\
&+\frac{1}{\epsilon ^3}\left(\frac{275}{9} \zeta _5-\frac{7033}{54} \zeta _3 \zeta
   _2\right)+\frac{1}{\epsilon ^2}\left(-\frac{210031}{648} \zeta _3^2-\frac{9349}{105} \zeta _2^3\right)+\mathcal{O}\left(\epsilon^{-1}\right)
\\
I_{7}^{(26)} &= \frac{1}{\epsilon^8}\left(\frac{1}{288}\right)+\frac{1}{\epsilon ^6}\left(\frac{1}{144} \zeta _2\right)+\frac{1}{\epsilon ^5}\left(\frac{209}{216} \zeta _3\right)+\frac{1}{\epsilon ^4}\left(\frac{43}{40} \zeta _2^2\right)
\nonumber \\
&+\frac{1}{\epsilon ^3}\left(-\frac{5761}{360} \zeta
   _5+\frac{59}{27} \zeta _3 \zeta _2\right)+\frac{1}{\epsilon ^2}\left(\frac{27179}{648} \zeta _3^2-\frac{17501}{2520} \zeta _2^3\right)+\mathcal{O}\left(\epsilon^{-1}\right)
\\
I_{8}^{(27)} &= \frac{1}{\epsilon ^8}\left(-\frac{1}{64}\right)+\frac{1}{\epsilon ^6}\left(\frac{5}{24} \zeta _2\right)+\frac{1}{\epsilon ^5}\left(\frac{55}{48} \zeta _3\right)+\frac{1}{\epsilon ^4}\left(\frac{49}{320} \zeta _2^2\right)
\nonumber \\
&+\frac{1}{\epsilon ^3}\left(\frac{1183}{240} \zeta _5-\frac{3395}{288} \zeta _3 \zeta
   _2\right)+\frac{1}{\epsilon ^2}\left(-\frac{20561}{576}\zeta _3^2-\frac{39377}{1680} \zeta _2^3\right)+\mathcal{O}\left(\epsilon^{-1}\right)
\\
I_{9}^{(28)} &= \frac{1}{\epsilon ^8}\left(-\frac{1}{96}\right)+\frac{1}{\epsilon ^6}\left(\frac{97}{288} \zeta _2\right)+\frac{1}{\epsilon ^5}\left(\frac{271}{144} \zeta _3\right)+\frac{1}{\epsilon ^4}\left(-\frac{3793}{2880} \zeta _2^2\right)
\nonumber \\
&+\frac{1}{\epsilon ^3}\left(\frac{4291}{120} \zeta _5-\frac{21359}{432} \zeta _3 \zeta
   _2\right)+\frac{1}{\epsilon ^2}\left(-\frac{19235}{144} \zeta _3^2+\frac{1397}{576} \zeta _2^3\right)+\mathcal{O}\left(\epsilon^{-1}\right)
\\
I_{10}^{(29)} &= \frac{1}{\epsilon^8}\left(-\frac{1}{1152}\right)+\frac{1}{\epsilon ^6}\left(-\frac{1}{576} \zeta _2\right)+\frac{1}{\epsilon ^5}\left(-\frac{13}{1728} \zeta _3\right)+\frac{1}{\epsilon ^4}\left(\frac{169}{640} \zeta _2^2\right)
\nonumber \\
&+\frac{1}{\epsilon ^3}\left(\frac{26357}{2880} \zeta _5+\frac{685}{1728} \zeta _3 \zeta _2\right)+\frac{1}{\epsilon ^2}\left(\frac{186637}{10368} \zeta _3^2+\frac{57191}{5040} \zeta _2^3\right)+\mathcal{O}\left(\epsilon^{-1}\right)
\\
I_{11}^{(30)} &= I_{\mathrm{p}, 20}^{(30)}
\\
I_{12}^{(27)} &= \frac{1}{\epsilon ^8}\left(\frac{35}{1152}\right)+\frac{1}{\epsilon ^6}\left(-\frac{73}{192} \zeta _2\right)+\frac{1}{\epsilon ^5}\left(-\frac{1015}{432} \zeta _3\right)+\frac{1}{\epsilon ^4}\left(-\frac{4069}{1440} \zeta _2^2\right)
\nonumber \\
&+\frac{1}{\epsilon ^3}\left(-\frac{8693}{144} \zeta
   _5+\frac{5809}{288} \zeta _3 \zeta _2\right)+\frac{1}{\epsilon ^2}\left(\frac{260783}{5184} \zeta _3^2-\frac{36499}{2240}\zeta _2^3\right)+\mathcal{O}\left(\epsilon^{-1}\right)
\\
I_{13}^{(28)} &= \frac{1}{\epsilon ^8}\left(-\frac{13}{1152}\right)+\frac{1}{\epsilon ^6}\left(\frac{35}{192} \zeta _2\right)+\frac{1}{\epsilon ^5}\left(\frac{305}{432} \zeta _3\right)+\frac{1}{\epsilon ^4}\left(\frac{461}{1440} \zeta _2^2\right)
\nonumber \\
&+\frac{1}{\epsilon ^3}\left(\frac{4001}{180} \zeta _5-\frac{461}{72}\zeta _3 \zeta
   _2\right)+\frac{1}{\epsilon ^2}\left(\frac{11243}{324} \zeta _3^2+\frac{4295}{336} \zeta _2^3\right)+\mathcal{O}\left(\epsilon^{-1}\right)
\\
I_{14}^{(29)} &= \frac{1}{\epsilon ^5}\left(-\frac{1}{32}\zeta _3\right)+\frac{1}{\epsilon ^4}\left(\frac{9}{320} \zeta _2^2\right)+\frac{1}{\epsilon ^3}\left(-\frac{371}{96} \zeta _5+\frac{91}{48} \zeta _3 \zeta _2\right)
\nonumber \\
&+\frac{1}{\epsilon ^2}\left(-\frac{223}{96} \zeta _3^2+\frac{653}{576} \zeta
   _2^3\right)+\mathcal{O}\left(\epsilon^{-1}\right)
\\
I_{15}^{(29)} &= \frac{1}{\epsilon ^8}\left(-\frac{1}{18}\right)+\frac{1}{\epsilon ^6}\left(\frac{53}{48} \zeta _2\right)+\frac{1}{\epsilon ^5}\left(\frac{2621}{432} \zeta _3\right)+\frac{1}{\epsilon ^4}\left(-\frac{1423}{1440} \zeta _2^2\right)
\nonumber \\
&+\frac{1}{\epsilon ^3}\left(\frac{54437}{720} \zeta _5-\frac{7751}{72} \zeta _3 \zeta
   _2\right)+\frac{1}{\epsilon ^2}\left(-\frac{413683}{1296} \zeta _3^2-\frac{410153}{10080} \zeta _2^3\right)+\mathcal{O}\left(\epsilon^{-1}\right)
\\
I_{16}^{(30)} &= \frac{1}{\epsilon ^8}\left(\frac{7}{192}\right)+\frac{1}{\epsilon ^6}\left(-\frac{35}{96} \zeta _2\right)+\frac{1}{\epsilon ^5}\left(-\frac{271}{144} \zeta _3\right)+\frac{1}{\epsilon ^4}\left(-\frac{49}{160} \zeta _2^2\right)
\nonumber \\
&+\frac{1}{\epsilon ^3}\left(-\frac{6037}{240} \zeta
   _5+\frac{3343}{144} \zeta _3 \zeta _2\right)+\frac{1}{\epsilon ^2}\left(\frac{42271}{864} \zeta _3^2+\frac{1711}{315} \zeta _2^3\right)+\mathcal{O}\left(\epsilon^{-1}\right)
\\
I_{17}^{(30)} &= \frac{1}{\epsilon ^5}\left(-\frac{1}{32}\zeta _3\right)+\frac{1}{\epsilon ^4}\left(\frac{37}{960} \zeta _2^2\right)+\frac{1}{\epsilon ^3}\left(\frac{49}{96} \zeta _5+\frac{9}{16} \zeta _3 \zeta _2\right)
\nonumber \\
&+\frac{1}{\epsilon ^2}\left(-\frac{625}{96} \zeta _3^2+\frac{81401}{20160}\zeta
   _2^3\right)+\mathcal{O}\left(\epsilon^{-1}\right)
\\
I_{18}^{(30)} &= \frac{1}{\epsilon ^8}\left(\frac{1}{288}\right)+\frac{1}{\epsilon ^6}\left(\frac{11}{288} \zeta _2\right)+\frac{1}{\epsilon ^5}\left(-\frac{1}{864}\zeta _3\right)+\frac{1}{\epsilon ^4}\left(-\frac{241}{160} \zeta _2^2\right)
\nonumber \\
&+\frac{1}{\epsilon ^3}\left(-\frac{18559}{1440} \zeta
   _5-\frac{9749}{864} \zeta _3 \zeta _2\right)+\frac{1}{\epsilon ^2}\left(-\frac{153467}{5184}\zeta _3^2-\frac{763019}{20160}\zeta _2^3\right)+\mathcal{O}\left(\epsilon^{-1}\right)
\\
I_{19}^{(22)} &= \frac{1}{\epsilon ^8}\left(\frac{1}{576}\right)+\frac{1}{\epsilon ^6}\left(\frac{29}{288} \zeta _2\right)+\frac{1}{\epsilon ^5}\left(\frac{269}{432} \zeta _3\right)+\frac{1}{\epsilon ^4}\left(\frac{553}{360} \zeta _2^2\right)
\nonumber \\
&+\frac{1}{\epsilon ^3}\left(\frac{43109}{720} \zeta _5-\frac{349}{27} \zeta _3 \zeta
   _2\right)+\frac{1}{\epsilon ^2}\left(\frac{57485}{1296} \zeta _3^2+\frac{142267}{2520} \zeta _2^3\right)+\mathcal{O}\left(\epsilon^{-1}\right)
\\
I_{20}^{(22)} &= \frac{1}{\epsilon ^2}\left(\frac{1}{4}\zeta_3^2+\frac{31}{140} \zeta _2^3\right)+\mathcal{O}\left(\epsilon^{-1}\right)
\\
I_{21}^{(24)} &= \frac{1}{\epsilon ^8}\left(\frac{5}{576}\right)+\frac{1}{\epsilon ^6}\left(\frac{37}{288} \zeta _2\right)+\frac{1}{\epsilon ^5}\left(\frac{229}{432} \zeta _3\right)+\frac{1}{\epsilon ^4}\left(-\frac{541}{360} \zeta _2^2\right)
\nonumber \\
&+\frac{1}{\epsilon ^3}\left(\frac{1799}{144} \zeta _5-\frac{13385}{432} \zeta _3 \zeta
   _2\right)+\frac{1}{\epsilon ^2}\left(-\frac{259405}{2592}\zeta _3^2-\frac{222371}{10080} \zeta _2^3\right)+\mathcal{O}\left(\epsilon^{-1}\right)
\\
I_{22}^{(24)} &= \frac{1}{\epsilon ^8}\left(\frac{1}{144}\right)+\frac{1}{\epsilon ^6}\left(-\frac{1}{18}\zeta _2\right)+\frac{1}{\epsilon ^5}\left(-\frac{263}{216} \zeta _3\right)+\frac{1}{\epsilon ^4}\left(-\frac{3127}{720} \zeta _2^2\right)
\nonumber \\
&+\frac{1}{\epsilon ^3}\left(-\frac{27287}{360} \zeta
   _5+\frac{683}{108} \zeta _3 \zeta _2\right)+\frac{1}{\epsilon ^2}\left(\frac{35743}{648} \zeta _3^2-\frac{71705}{1008} \zeta _2^3\right)+\mathcal{O}\left(\epsilon^{-1}\right)
\\
I_{23}^{(28)} &= \frac{1}{\epsilon ^4}\left(\frac{1}{4}\zeta _2^2\right)+\frac{1}{\epsilon ^3}\left(\frac{5}{4} \zeta _5+\frac{3}{4} \zeta _3 \zeta _2\right)+\frac{1}{\epsilon ^2}\left(\frac{117}{8} \zeta _3^2-\frac{659}{168} \zeta _2^3\right)
+\mathcal{O}\left(\epsilon^{-1}\right)
\end{align}
The above expressions were derived from results for finite master integrals calculated in reference \cite{paperQCDcusp} for the determination of the cusp anomalous dimensions of massless QCD.
A key feature of the finite integral analysis is that the most complicated, non-linearly reducible master integrals, {\it e.g.} $I_{\mathrm{p}, 16}^{(19)}$, $I_{\mathrm{p}, 17}^{(19)}$, $I_{6}^{(26)}$, and $I_{7}^{(26)}$, may be computed through to weight six by judiciously choosing finite integrals in the relevant integral topologies which first contribute to the form factor at transcendental weight seven ({\it i.e.} at the level of the $\epsilon^{-1}$ pole).
The finite integrals were defined allowing for shifted dimensions and additional powers of the propagators (dots) \cite{vonManteuffel:2014qoa,vonManteuffel:2015gxa,Panzer:2014gra,Schabinger:2018dyi} using the integral finder in \Reduze~\cite{vonManteuffel:2012np} and, in linearly-reducible cases, integrated in the Feynman parametric representation using \HyperInt~\cite{Panzer:2014caa}.
In order to express the above 55 integrals from reference \cite{Boels:2017ftb} in terms of finite integrals,
linear relations between integrals were computed using finite field arithmetic~\cite{vonManteuffel:2014ixa,vonManteuffel:2016xki}. Moreover, syzygies were employed to avoid numerators in the reductions of integrals with many dots \cite{Lee:2014tja,Bitoun:2017nre,vonManteuffel:2019wbj}.
Note that analytic results for a subset of the integrals discussed here were presented in previous works~\cite{Henn:2016men,Lee:2016ixa,Lee:2019zop,vonManteuffel:2019wbj,vonManteuffel:2019gpr}.

\section{Results}
\label{sec:results}
\subsection{The ${\mathcal{N}=4}$ Four-Loop Sudakov Form Factor to Weight Six}
\label{sec:ffresult}
Combining the formulae contained in Sections \ref{sec:ffintegrand} and \ref{sec:ffmasters}, we find
\begin{align}
\label{eq:ffexp}
F^{(4)} &= \frac{1}{\epsilon ^8}\left(\frac{2}{3}\right)+\frac{1}{\epsilon ^6}\left(\frac{2}{3}\zeta _2\right)+\frac{1}{\epsilon ^5}\left(-\frac{38}{9} \zeta _3\right)+\frac{1}{\epsilon ^4}\left(\frac{5}{18}\zeta _2^2\right)+\frac{1}{\epsilon ^3}\left(\frac{1082}{15} \zeta
   _5+\frac{23}{3} \zeta _3 \zeta _2\right)
\nonumber \\
&
   +\frac{1}{\epsilon ^2}\left(\frac{10853}{54} \zeta_3^2+\frac{95477}{945} \zeta _2^3\right)
+\frac{1}{N_c^2}\bigg[\frac{1}{\epsilon^2}\left(18 \zeta _3^2+\frac{372}{35} \zeta _2^3\right)\bigg]+\mathcal{O}\left(\epsilon^{-1}\right).
\end{align}
It is worth pointing out that, via the principle of maximal transcendentality \cite{Kotikov:2001sc,Kotikov:2002ab}, the $\epsilon^{-8}-\epsilon^{-3}$ poles may be inferred in a straightforward manner from the renormalization group predictions of reference \cite{Moch:2005id} for the four-loop quark form factor of massless QCD.
Alternatively, they can be predicted by the requirement that the logarithm of the form factor has at most a double pole in $\epsilon$.

\subsection{The ${\mathcal{N} = 4}$ Four-Loop Cusp Anomalous Dimension}
\label{sec:cusp}
The renormalization group analysis of reference \cite{Moch:2005id}, together with the known higher-order-in-$\epsilon$ results for the one-, two-, and three-loop form factors \cite{Gehrmann:2011xn}, we see that the $\epsilon^{-2}$ pole of the $\mathcal{N} = 4$ Sudakov form factor must be
\begin{align}
-\frac{1}{32}\Gamma_4^{\mathcal{N} = 4}+\frac{10799}{54} \zeta _3^2+\frac{89564}{945} \zeta _2^3,    
\end{align}
where $\Gamma_4^{\mathcal{N} = 4}$ is the four-loop cusp anomalous dimension of the $\mathcal{N}=4$ model. Its relation to $\gamma_{\rm cusp}^{(4)}$ defined in \cite{Boels:2017skl,Boels:2017ftb} is $\gamma_{\rm cusp}^{(4)} = 2 \, \Gamma_4^{\mathcal{N} = 4}$. By comparing to Eq.\ \eqref{eq:ffexp} above, we see immediately that
\begin{align}
\label{eq:cusp}
\Gamma_4^{\mathcal{N} = 4} = -32 \zeta _3^2-\frac{7008}{35} \zeta _2^3 + \frac{1}{N_c^2}\bigg[-576 \zeta _3^2-\frac{11904}{35} \zeta _2^3\bigg].
\end{align}
While the leading-color part of Eq. \eqref{eq:cusp} has long been known \cite{Bern:2006ew, Beisert:2006ez,Henn:2013wfa}, our calculation of $\Gamma_4^{\mathcal{N} = 4}$ provides a strong check on the original numerical analysis of the $\mathcal{N} = 4$ form factor \cite{Boels:2017skl,Boels:2017ftb} and on the analytic four-loop $\mathcal{N} = 4$ Wilson loop analysis in \cite{Henn:2019swt}.

\section{Conclusions}
\label{sec:conclusions}
We calculated the full four-loop cusp anomalous dimension of $\mathcal{N}=4$ supersymmetric Yang-Mills theory analytically.
Our result was derived from the four-loop Sudakov form factor using parametric integrations of finite master integrals calculated in \cite{paperQCDcusp} for the determination of the cusp anomalous dimensions in QCD.
In our approach, the most complicated integral topologies decouple from the calculation of the cusp because their finite master integrals may be judiciously selected to first contribute to the $\epsilon^{-1}$ pole of the Sudakov form factor. Our calculation confirms the result of the very recent independent calculation of the $\mathcal{N}=4$ cusp anomalous dimension in \cite{Henn:2019swt} based on the Wilson loop picture.
Our findings are in agreement with the earlier semi-numerical analysis of \cite{Boels:2017skl,Boels:2017ftb} at the level of the master integrals, for which we provide uniformly transcendental analytic results through to weight six.
The analytic results for the master integrals strongly suggest that the full four-loop form factor in $\mathcal{N}=4$ super Yang-Mills is uniformly transcendental.

\section*{Acknowledgments}
We would like to thank Rutger Boels for helpful discussions and collaboration at an early stage of this work. We also thank Nils Kreher for useful correspondence. This work employed computing resources provided by the
High Performance Computing Center at Michigan State University,
and we gratefully acknowledge the HPCC team for their help and support.
Computing resources at the supercomputer Mogon at Johannes Gutenberg University Mainz were employed in addition,
and we would like to thank Hubert Spiesberger, the PRISMA excellence cluster, and the Mogon team for their generous support.
We also thank the mathematical institutes of the University of Oxford and Humboldt University of Berlin for computing resources.
The work of TH was supported in part by the Deutsche Forschungsgemeinschaft (DFG, German Research Foundation) under grant  396021762 - TRR 257 ``Particle Physics Phenomenology after the Higgs Discovery.''
The work of AvM was supported in part by the National Science Foundation under Grant No.\ 1719863.
The work of GY was supported in part by the National Natural Science Foundation of China (Grants No. 11822508, 11847612, 11935013) and by the Key Research Program of Frontier Sciences of Chinese Academy of Sciences.

\bibliographystyle{JHEP}
\bibliography{neq4cusp}

\end{document}